\documentclass[12pt]{iopart}

\usepackage{graphicx}
\usepackage{bbold}
 
\begin{document}

\title[]{Extracting atoms one by one from a small matter-wave soliton}

\author{Fatema~Hamodi and Lev~Khaykovich}

\address{Department of Physics, QUEST Center and Institute of Nanotechnology and Advanced Materials, Bar-Ilan University, Ramat Gan, 5290002, Israel}
\vspace{10pt}
\begin{indented}
\item[]August 2019
\end{indented}

\begin{abstract}
Excitations of small one-dimensional matter-wave solitons are considered within a framework of the attractive Bose-Hubbard model.
The initial eigenstates of the system are found by exact diagonalization of the Bose-Hubbard Hamiltonian.  
We drive transitions between the eigenstates by inducing a weak modulation of the tunnelling rate and show that a single atom can be extracted while the remaining atoms stay localized despite the persistent external modulation.
This scheme suggests the experimental realization of small matter-wave solitons with deterministic number of atoms.
In addition, the knowledge of exact eigenstates allows identification of the selection rules for transitions between the different eigenstates of the Hamiltonian.
One selection rule is related to the translation symmetry of the system.
Another one is strictly applicable only on a subspace of the total Hilbert space and is related to the parity symmetry.
We show that in the strongly interacting limit this selection rule has implications on the entire Hilbert space.
We discuss its signatures on the system's dynamics and consider how it can be observed experimentally with ultracold atoms.
\end{abstract}

%
%
%
%
%

\section{Introduction}
In recent years, experimental abilities to prepare well defined states with a deterministic number of atoms reached a new level of precision.
In few-fermion systems a clever combination of the Pauli exclusion principle and an external harmonic confinement allowed the researchers to prepare such states in a single optical dipole trap~\cite{Serwane11}.
For bosonic systems, a relatively complex manipulation of many individual atom traps is required to achieve the goal~\cite{Barredo16,Endres16,Kim16}.
Here we consider theoretically a relatively simple protocol applied to 1D bosonic samples with attractive interactions as an alternative avenue for preparing deterministic few-boson states in a single trap.

A 1D attractive Bose gas supports a solitonic solution known as a bright soliton~\cite{PS16}.
The phase transition toward this translational symmetry breaking solution has been a subject of theoretical research~\cite{Kanamoto03,Zin08} mainly within the framework of the mean-field approach, i.e. Gross-Pitaevskii and Bogoliubov theories.
All recent experiments related to bright solitons~\cite{Marchant13,Marchant16,Nguyen14,Nguyen17,McDonald14,Boisse17,Wales19} have been performed in the regime where the mean-field approximation is valid.

In this limit, the degrees of freedom of the relative motion of the atoms within the soliton and the center-of-mass (CoM) motion of the soliton as a whole are unseparable.
In contrast, the full quantum mechanical treatment separates them and leads to the investigation of fundamental quantum mechanical properties of solitons~\cite{Weiss09,Streltsov09,Gertjerenken12,Gertjerenken13} and their possible applications in future quantum devices~\cite{Helm14}.
For example, Refs.~\cite{Weiss09,Streltsov09} predict the formation of quantum superposition states through soliton scattering off a potential barrier based precisely on this separation.

The beyond mean-filed approach can be conveniently studied within the framework of the Bose-Hubbard model (BHM).
The presence of an external periodic potential, required by the model, enriches the initial problem and leads to interesting consequences, one of which is considered here.  
The phase transition and some properties of the ground state of the attractive BHM have been studied in Refs.~\cite{Oelkers07,Jack05,Sorensen12}, and more on static and dynamic analysis of this model can be found in Refs. \cite{Barbiero14,Naldesi18}.

In this paper we consider induced transitions in a few-boson system with attractive interactions.
The atoms are loaded into a one-dimensional optical lattice in the tight binding regime where the BHM is applicable.
We consider a finite optical lattice with periodic boundary conditions and apply an exact numerical diagonalization method to find the energy spectrum and the eigenstates of the system.
We then induce resonance transitions between different energy states by the introduction of a weak modulation of the tunnelling rate in the Hamiltonian. 
We solve a system of coupled Shr{\"o}dinger equations by direct integration and show that a single atom can be extracted from the solitonic state.
We demonstrate two consecutive steps of this scheme.
Direct extension of the model suggests the possibility of cascading extraction of atoms from the initial solitonic state one by one and preparation of small matter-wave solitons with determenistic number of atoms.  

The knowledge of the exact eigenstates of the problem allows the identification of the selection rules for the induced transitions.
One obvious selection rule is related to the translational symmetry of the problem and reflects the conservation of quasi-momentum.
However, we identify another selection rule which applies for a certain subspace of the total Hilbert space and is related to the parity symmetry.
We show that this selection rule, although strictly applicable only on the zero quasi-momentum subspace, has much wider implications on the system due to the solitonic character of the eigenstates.
We show how these selection rules affect population probabilities of the induced transitions and suggest possible experimental verification of the effect using ultracold atoms.

\section{Bose-Hubbard model}
\label{sec:BHM}
\subsection{Stationary Hamiltonian}
\label{sec:BHMSH}
We consider $N$ particles distributed on a 1-D optical lattice with $M$ sites and periodic boundary conditions.
In the tight binding approximation the system is described by the Bose-Hubbard Hamiltonian:
\begin{equation}
\hat{H}_0=-J\sum_j({\hat{a}_{j+1}^\dagger}\hat{a}_j + H.c) +\frac{U}{2}\sum_j\hat{n}_j(\hat{n}_j-1),
\label{Eq:Hamiltonian}
\end{equation}
where $J$ is the tunnelling strength and $U$ is the on site interaction strength. 
In case of attractive interactions $U<0$.
The operator $\hat{a}_j^\dagger$ ($\hat{a}_j$) creates (annihilates) a particle on site $j$ of the lattice.
The usual method to work with this Hamiltonian is to use the Fock state basis $|n_0,n_1,...,n_{M-1}\rangle$, where the total number of atoms is $N=n_0+n_1+...+n_{M-1}$. 

We then define the translation operator $\hat{\mathcal{T}}$:
\begin{equation}
\label{Eq:TranslationOp}
\hat{\mathcal{T}}|n_0,n_1,...,n_{M-1}\rangle\equiv|n_{M-1},n_0,n_1,...,n_{M-2}\rangle,
\end{equation} 
and note that $\hat{\mathcal{T}}^M = \mathbb{1}$. 
Thus the eigenvalues of the translational operator have to be the $M$th root of unity.
In the simplest form, this can be expressed as 
\begin{equation}
\hat{\mathcal{T}}|\Psi_{n}\rangle = e^{-iq_n}|\Psi_{n}\rangle,
\label{Eq:TranslationOpEigenV}
\end{equation}
where  $q_n=2\pi n/M$ is the quasi-momentum of an eigenstate $|\Psi_{n}\rangle$ with $n$ being an integer.

\begin{figure}
\centering\includegraphics[width=0.4\columnwidth]{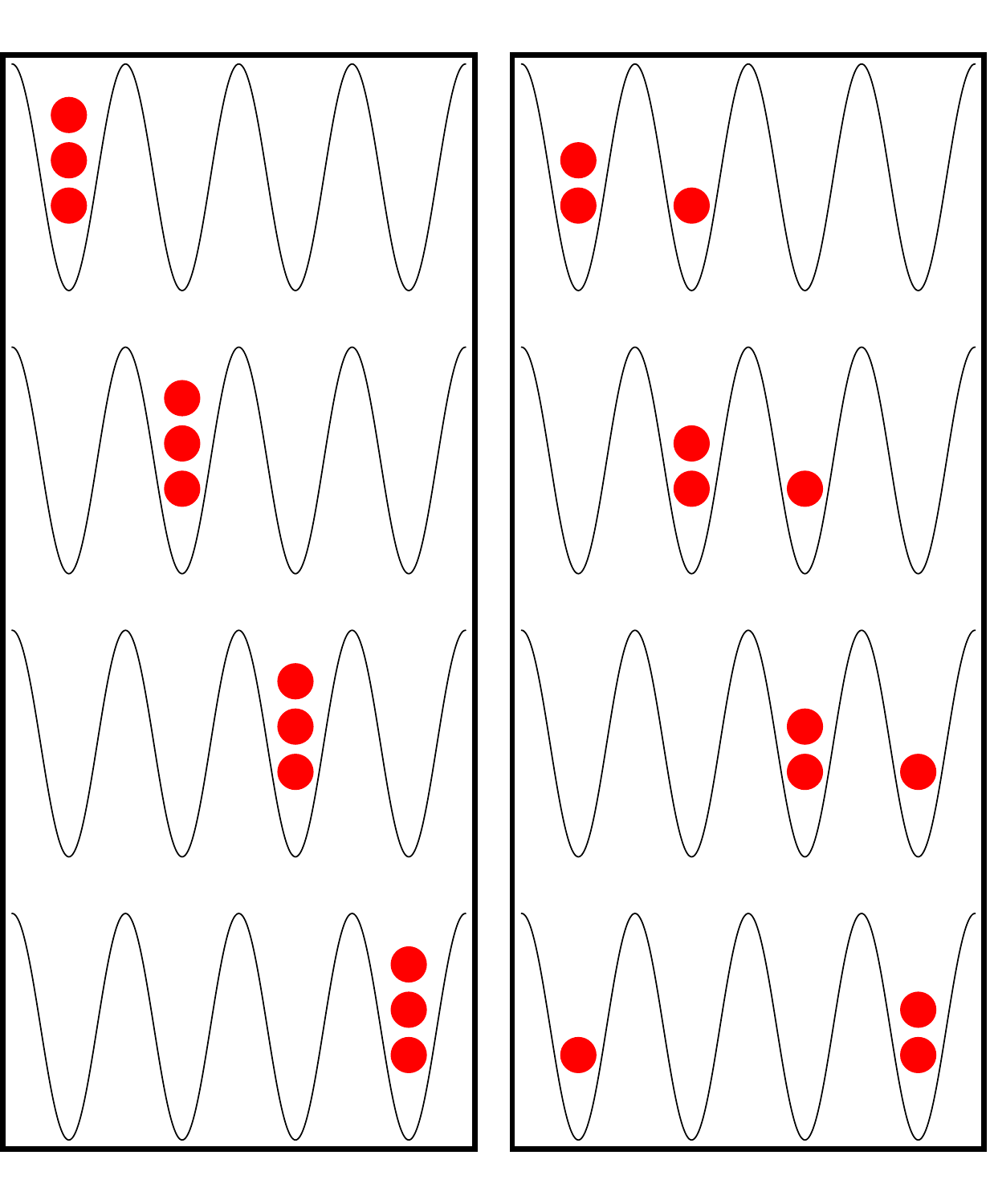}
\caption{\label{fig:EquivalanceClass} A schematic representation of two equivalence classes for the system of $N=3$ particles on $M=4$ sites. The left (right) class, consisting of all translational copies of the Fock state with all atoms occupying a single site (two atoms on one site and a single atom on the neighbouring site), is shown in the left (right) column. }
\end{figure}

Using the commutativity of the Hamiltonian~(\ref{Eq:Hamiltonian}) with the translation operator $\hat{\mathcal{T}}$, we apply the exact diagonalization method in the basis of the translation operator eigenstates to solve the eigenvalue problem.
Following the treatment of Ref.~\cite{Sorensen12}, we divide the basis of Fock states into $P$ equivalence classes: 
\begin{equation}
\mathcal{E}^{(j)}=\{|\Phi_0^{(j)}\rangle,|\Phi_1^{(j)}\rangle,...,|\Phi_{M-1}^{(j)}\rangle\},
\end{equation}
with $j=0,1,..,P-1$, such that the two Fock states $|\Phi_a\rangle$ and $|\Phi_b\rangle$ belong to the same equivalence class if $|\Phi_b\rangle=\hat{\mathcal{T}}^k|\Phi_a\rangle$ for some integer $k$.
To illustrate this idea we show two examples of such equivalence classes in Fig.~\ref{fig:EquivalanceClass} for a specific system of $N=3$ particles on $M=4$ sites.
This first class includes all translation copies of the Fock states with all 3 particles occupying the same site (left column) while the second class covers translation copies of two occupied neigbouring sites: the first one with 2 atoms and the next one with 1 atom (right column).

Then we choose the basis:
\begin{equation}
\label{Eq:basis}
|\xi^{(j)}_n\rangle = \sqrt{\frac{1}{M}}\sum_{k=0}^{M-1} e^{-ikq_n}|\Phi_k^{(j)}\rangle,
\end{equation}
where $n=0,1,...,M-1$.
Eq.~(\ref{Eq:basis}) is the eigenstate of the translation operator with the eigenvalue shown in Eq.~(\ref{Eq:TranslationOpEigenV}).
In this basis the representation of $\hat{H}$ becomes block-diagonal which simplifies the original diagonalizaiton problem.

\begin{figure}
\centering\includegraphics[width=0.6\columnwidth]{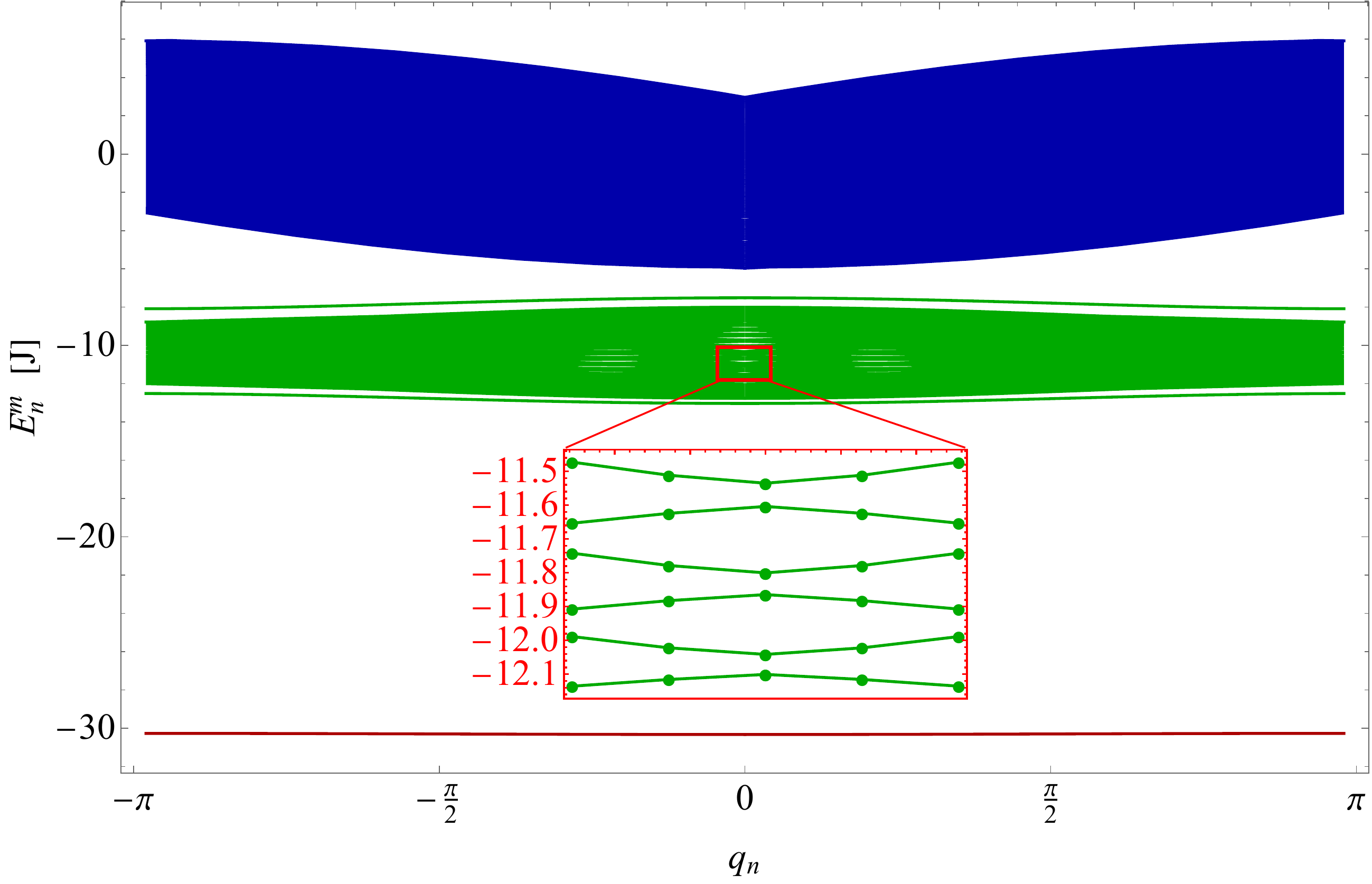}
\caption{\label{fig:EnergySpectrum} Energy diagram of $N=3$ particles on $M=149$ sites for $U/J=-10$. The inset emphasises the discrete structure of the spectra. The ground state (red line) corresponds to all $3$ atoms in the localized state. The first excited (green) band describes $2$ atoms in the localized state and $N_f=1$ free atom. We exclude from the discussion the lowest and the highest levels in this band which has a different character as discussed in Ref.~\cite{Valiente10}. The second excited (blue) band corresponds to $N_f=3$ free atoms.}
\end{figure}

In Fig.~\ref{fig:EnergySpectrum} the energy spectra of $N=3$ particles and $M=149$ sites is shown for $U/J=-10$. 
In this limit of strong interactions the energy spectrum is split into energy bands which can be associated with different modes of atomic motion~\cite{Sorensen12}. 
An energy level in the spectrum corresponds to the CoM motion of the localized many-body state as a whole.
Close energy levels within the band describe the different modes of motion of atoms excited out of the solitonic state which we call "free" atoms.
Of course, these "free" atoms still see the periodic potential.
The band gap between two consecutive bands corresponds to the differences in the relative motions of $N-N_f$ atoms in the localized state and $N_f$ free atoms and the relative motion of $N-N_f-1$ atoms in the localized state and $N_f+1$ free atoms.
In the specific example shown in Fig.~\ref{fig:EnergySpectrum}, the ground (red line) level describes the CoM motion of the localized state of 3 particles.
The first excited band corresponds to 2 atoms in the bound state and each individual level describes a mode of translational motion of the free atom. 
The third band corresponds to the disintegrated state of 3 free atoms.
For more detailed discussion of the energy spectrum structure we refer the reader to Ref.~\cite{Sorensen12}.

\subsection{The time dependent model}
To induce transitions between the different bands of the energy spectrum we introduce a time dependent perturbation of the tunnelling rate in the Hamiltonian~(\ref{Eq:Hamiltonian}):
\begin{eqnarray}
\hat{H}(t) = \hat{H}_0 + \hat{J}(t) = \hat{H}_0+\varepsilon \sin(\omega t)\hat{H}_k,  \nonumber\\
\mathrm{where} \qquad \hat{H}_k = -J\sum_j({\hat{a}_{j+1}^\dagger}\hat{a}_j + H.c).
\label{Eq:TDHamiltonian}
\end{eqnarray}
In this equation $\varepsilon\ll 1$ and $\omega$ denote the relative amplitude and frequency of the modulation of the tunnelling rate, respectively.
The modulation frequency is tuned to resonance with the energy difference of two consecutive bands.
Specifically, throughout this paper we choose the initial state of the system as the ground state and we couple it to an energy level in the first excited band with the energy difference: 
\begin{equation}
\label{Eq:EnergyDifference}
\omega = \frac{U}{2}\left(N(N-1)-(N-1)(N-2)\right).
\end{equation}
This difference matches the resonance condition between the ground state and an excited state for which the transition amplitude is maximal (see Fig.~\ref{fig:MatrixElements}(b)).

Now, any state of the system can be expanded in the basis~(\ref{Eq:basis}) and the time dependent Schr{\"o}dinger equation can be solved to study the transition rates.
First, however, we turn to identify the selection rules that apply to these transitions under the action of the time dependent operator $\hat{J}(t)$.

\section{Symmetries and selection rules}
\label{sec:SelectionRules}
\subsection{Mirror operator}
An obvious selection rule, easily identified in the system, reflects the conservation of quasi-momentum derived from the translational symmetry of the system, i.e. transitions can be made only between the states with the same quasi-momentum (over vertical lines in Fig.~\ref{fig:EnergySpectrum}).

There is, however, another selection rule imposed on the system, which is related to a mirror (parity) symmetry.
To explain this symmetry, we define a mirror operator which swaps all the particle occupation numbers with respect to some center of the finite lattice returning the "mirror image" state:
\begin{equation}
\label{Eq:MirrorOp}
\hat{\mathcal{M}}|n_0,n_1,...,n_{M-1}\rangle \equiv |n_{M-1},n_{M-2},...,n_{0}\rangle
\end{equation} 
Applied twice, the mirror operator returns the original state, thus $\hat{\mathcal{M}}^2=\mathbb{1}$ and the eigenvalues of $\hat{\mathcal{M}}$ are $\pm 1$.
It can be easily shown that the mirror operator commutes with the Hamiltonian~(\ref{Eq:Hamiltonian}) $[\hat{\mathcal{M}},\hat{H}]=0$ but does not commute with the translational operator $[\hat{\mathcal{M}},\hat{\mathcal{T}}]\neq0$.
Therefore, it is impossible to construct simultaneous eigenstates of all three operators. 
However, this statement is correct when the entire Hilbert space is considered.
Below, we identify a sub-space in which mirror symmetry is conserved giving rise to a new selection rule.

Observing Eq.~\ref{Eq:TranslationOpEigenV}, we note that when the Hilbert space is limited to the case of zero quasi-momentum ($q_0=0$), the eigenvalue of the translational operator simplifies to $1$.
Then, in this sub-space $\hat{\mathcal{T}}$ and $\hat{\mathcal{M}}$ commute:
\begin{eqnarray}
\label{Eq:MTCommutate}
\langle \Psi_n^{\prime}|[\hat{\mathcal{T}},\hat{\mathcal{M}}]|\Psi_n\rangle = \langle \Psi_n^{\prime}|\hat{\mathcal{T}}\hat{\mathcal{M}}|\Psi_n\rangle - \langle \Psi_n^{\prime}|\hat{\mathcal{M}}\hat{\mathcal{T}}|\Psi_n\rangle\nonumber\\ = \langle \Psi_n^{\prime}|\hat{\mathcal{M}}|\Psi_n\rangle - \langle \Psi_n^{\prime}|\hat{\mathcal{M}}|\Psi_n\rangle = 0,
\end{eqnarray}
and the mirror symmetry is conserved.
To further extend the discussion of mirror symmetry we define {\it mirrored} and {\it vain} equivalence classes which will be especially helpful later, when we show the appearance of the quasi-selection rules in the rest of the Hilbert space.

\subsection{{\it Mirrored} and {\it vain} equivalence classes}
There are two possible outcomes when the mirror operator is applied to all states belonging to the same equivalence class $\mathcal{E}^{(j)}$. 
In the first one, all the states from another equivalent class are obtained:
\begin{equation}
\label{Eq:PairedClasses}
\hat{\mathcal{M}}\{\mathcal{E}^{(j)}\}=\{\mathcal{E}^{(j^\prime)}\},\; j\neq j^\prime.
\end{equation}
This means that the mirror operator fully projects the equivalence class $\mathcal{E}^{(j)}$ to $\mathcal{E}^{(j^\prime)}$, i.e. every Fock state in the first equivalence class is mirrored to a state from the second equivalence class with one-to-one correspondence. 
We, thus, call these classes {\it mirrored}.
In the second outcome, only states from the same equivalence class are obtained:
\begin{equation}
\label{Eq:UnpairedClasses}
\hat{\mathcal{M}}\{\mathcal{E}^{(j)}\}=\{\mathcal{E}^{(j)}\}.
\end{equation}
We call such an equivalence class {\it vain} to reflect its narcissistic character.
Note that in Fig.~\ref{fig:EquivalanceClass}, the left column shows a {\it vain} equivalence class, while the right column describes one of the {\it mirrored} equivalence classes whose pair can be easily defined by applying the mirror operator on it. 

\subsection{Selection rules for zero quasi-momentum $q_0=0$}
\label{Sec:SelRulesZeroQuasiMomentum}
\subsubsection{{\it Mirrored} and {\it vain} basis states}
Now, let us concentrate on zero quasi-momentum ($q_0 = 0$) sub-space for which special properties exist.
Here, Eqs.~(\ref{Eq:PairedClasses},\ref{Eq:UnpairedClasses}) can be directly extended to the basis states themselves (instead of equivalence classes). Correspondingly,  we define {\it mirrored} and {\it vain} basis states as:
\begin{equation}
\label{Eq:PUnPBasisStates}
\begin{array}{lll}
\hat{\mathcal{M}}|\xi_0^{(j)}\rangle = |\xi_0^{(j^\prime)}\rangle,\; j\neq j^\prime\\
\hat{\mathcal{M}}|\xi_0^{(j)}\rangle = |\xi_0^{(j)}\rangle,
\end{array}
\end{equation}
respectively. Again, the above property is correct due to the fact that eigenvalue of the translation operator is $1$ in this sub-space.

\subsubsection{Mirroring property of the eigenstates}
\label{sec:MirroringProperty}

\begin{figure}
\centering\includegraphics[width=0.6\columnwidth]{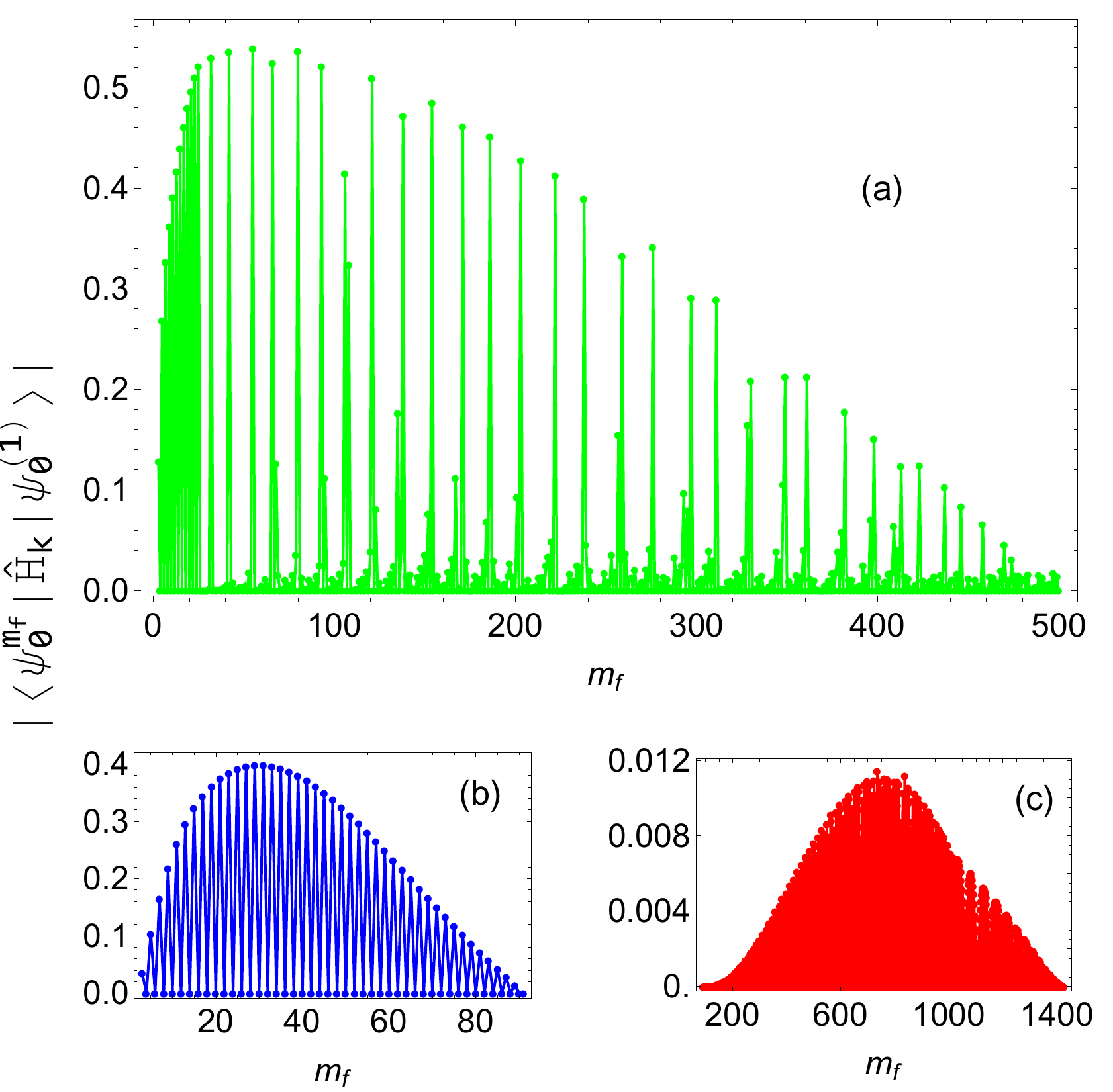}
\caption{\label{fig:MatrixElements} Numerical values of transition amplitudes for $q_{0}=0$ as a function of the final state number $m_f$ for $N=3$, $M=91$: (a) for $U/J = -3$ where there is no energy gap between the second and third excited bands; (b),(c) for $U/J=-10$ for energy levels from second and third excited bands respectively. The lines are guides to the eye. Note that $m_f=30$, which corresponds to maximum in transition amplitude in (b), satisfies the condition set by Eq.~(\ref{Eq:EnergyDifference}).}
\end{figure}

The eigenstates of the Hamiltonian~(\ref{Eq:Hamiltonian}) can be expanded in the chosen basis (\ref{Eq:basis}):
\begin{equation}
\label{Eq:EigenStateExpansion}
|\Psi_0^{(m)}\rangle = \sum_{j=1}^{P} C_j^{(m)} |\xi_0^{(j)}\rangle,
\end{equation}   
where $m$ is the excitation number and $C_j^{(m)}\equiv\langle \xi_0^{(j)}|\Psi_0^{(m)}\rangle$ are the overlap coefficients between the eigenstate and the basis states of the equivalence class $\mathcal{E}^{(j)}$.
Note, that these coefficients are real for $q_0$. 
In the general case of $M$ sites and $N$ particles, $m\in [1,...,P]$, where $P=\Gamma^M_N/M$ and 
\begin{equation}
\Gamma_N^M=\left(\begin{array}{cc} M+N-1\\
N
\end{array}\right).
\end{equation} 

We can now identify the rules $C_j^{(m)}$ are subject to in order to satisfy the requirement that $|\Psi_0^{(m)}\rangle$ are the eigenstates of the mirror operator $\hat{\mathcal{M}}$: 
\begin{equation}
\label{Eq:MirroringOfEigenstate}
\hat{\mathcal{M}}|\Psi_0^{(m)}\rangle = \sum_{j=0}^{P-1} C_j^{(m)} \hat{\mathcal{M}}|\xi_0^{(j)}\rangle 
=\sum_{j=0}^{P-1}  C_j^{(m)} |\xi_0^{(j^\prime)}\rangle,
\end{equation}
where $j^{\prime}$ is strictly related to $j$ according to Eq.~(\ref{Eq:PUnPBasisStates}).
We can now deduce the relation between $C_j^{(m)}$ and $C_{j^\prime}^{(m)}$ by requiring equality between Eq.~(\ref{Eq:MirroringOfEigenstate}) and the eigenstate expansion (\ref{Eq:EigenStateExpansion}), where $j$ is simply substituted by $j^{\prime}$. 
Recalling that eigenstates can be either odd or even, the following relations are obtained: for odd (even) mirroring eigenstates, $C_j^{(m)}=-C_{j^\prime}^{(m)} \; \& \; C_{vain}^{(m)}=0 \; \left( C_j^{(m)}=C_{j^\prime}^{(m)}\; \& \;C_{vain}^{(m)}\in\mathbb{R}\right)$ for {\it mirrored} $\&$ {\it vain} basis states. 
We verify that all coefficients $C_j^{(m)}$ obtained by the exact numerical diagonalization method indeed satisfy these relations.

To summarize, in the $q_{0}$ sub-space of the total Hilbert space, $\hat{\mathcal{M}}$ commutes with both $\hat{H}$ and $\hat{\mathcal{T}}$ and $|\Psi_0^{(m)}\rangle$ are the simultaneous eigenstates of all three operators.
As $|\Psi_0^{(m)}\rangle$ possesses an odd or even mirroring property, any induced transition has to preserve it.

The selection rule can be formulated as follows: the transition amplitudes $\langle \Psi_0^{m_f}|\hat{H}_k|\Psi_0^{m_i}\rangle$ vanish,
when the initial $|\Psi_0^{m_i}\rangle$ and final $|\Psi_0^{m_f}\rangle$ states are of the opposite mirroring, i.e. if $\langle\Psi_0^{m_f}|\hat{\mathcal{M}}|\Psi_0^{m_f}\rangle=-\langle\Psi_0^{m_i}|\hat{\mathcal{M}}|\Psi_0^{m_i}\rangle$.

In Fig.~\ref{fig:MatrixElements}, numerical values of the transition amplitudes from ground state to the final state $m_f$ are shown for the system of $N=3$ and $M=91$ and for two different interaction strengths: (a) $U/J=-3$ and (b),(c) $U/J=-10$.
For even states (the states which have opposite mirroring compared to the ground state) the transition amplitudes strictly vanish for any interaction strength.
The limit of strong interactions (Fig.~\ref{fig:MatrixElements}(b),(c)), whose energy diagram is represented in Fig.~\ref{fig:EnergySpectrum}, is of special interest for induced transitions discussed later (see Sec.~\ref{sec:InducedTransitions}).
Note, that we exclude transitions to the first ($m_f=2$) and the last excited states from our consideration. 
The special character of these sates is discussed in Ref.~\cite{Valiente10}.

\subsection{Quasi-selection rules for $q_n \neq 0$}
\begin{figure}
\centering\includegraphics[width=0.6\columnwidth]{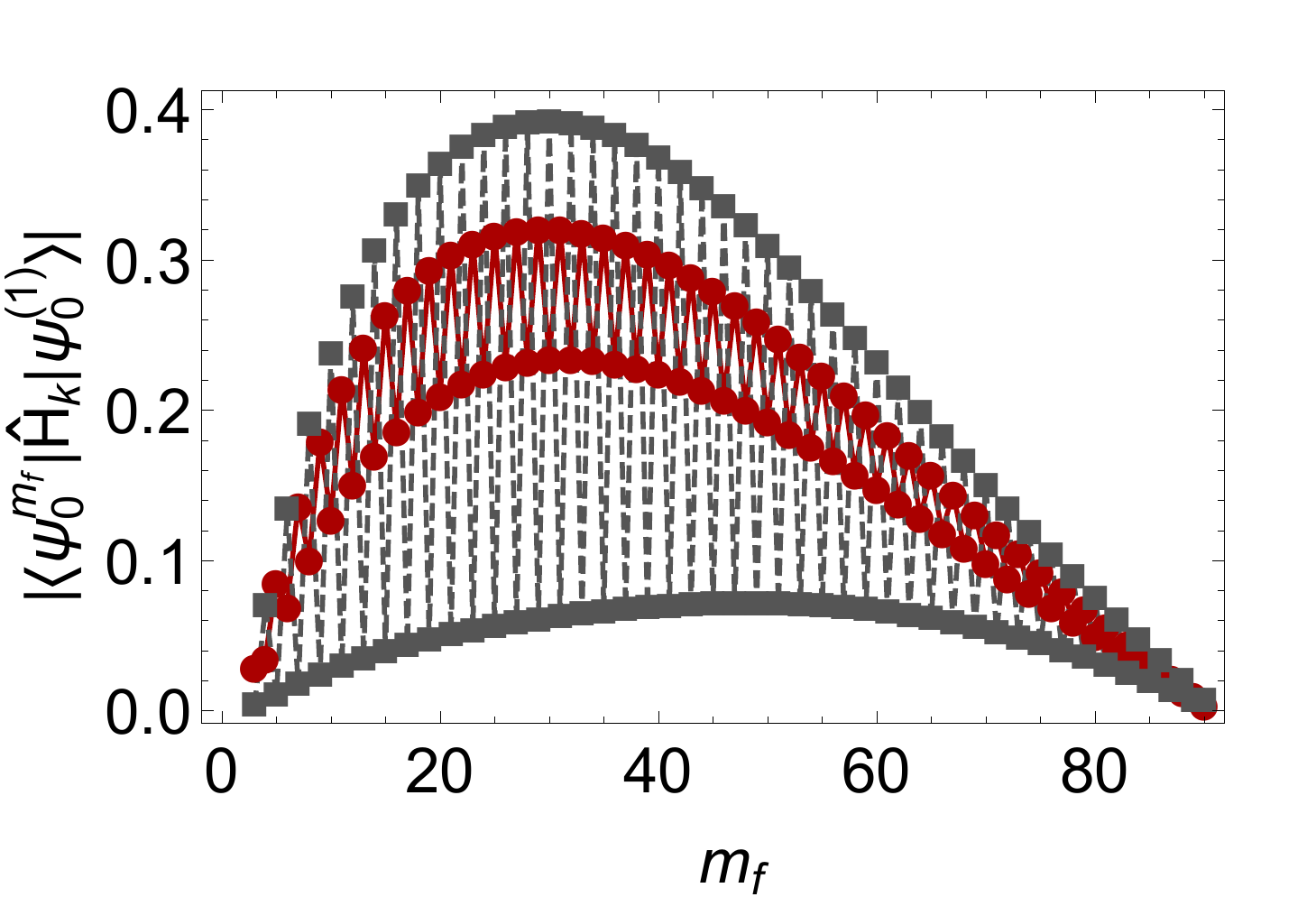}
\caption{\label{fig:MatrixElementsNZQM} Numerical values of transition amplitudes as a function of the final state number $m_f$ in the first excited band for $q_{n=1}$ (red circles) and $q_{n=3}$ (gray squares). Here we use $U/J=-10$, $N=3$ and $M=91$. The lines are guides to the eye.}
\end{figure}

\begin{figure}
\centering\includegraphics[width=0.6\columnwidth]{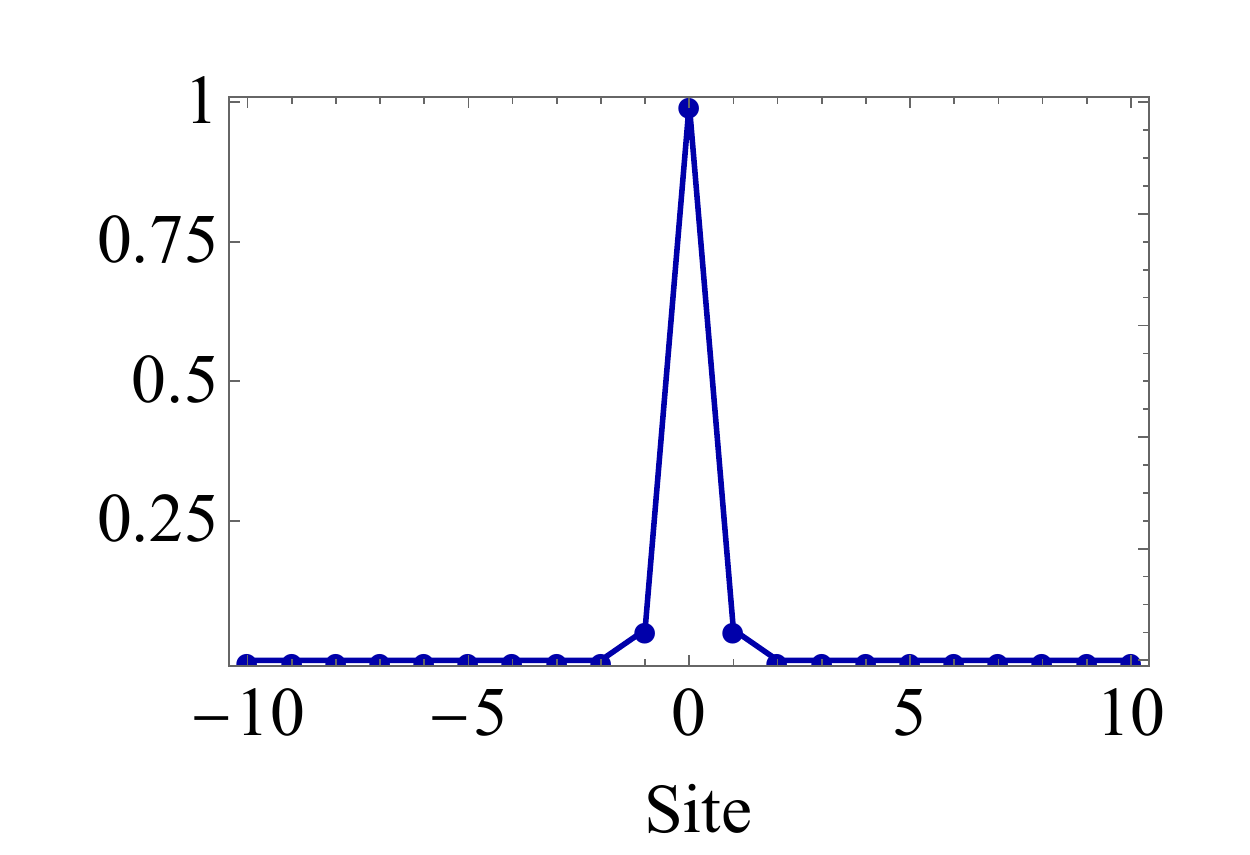}
\caption{\label{fig:SolitonicState} Wavefunction of a single localized state of the system of $N=3$ and $M=149$ in the strong interaction limit ($U/J=-10$). The full ground state wavefunction~(\ref{Eq:EigenStateExpansion}) consists of a superposition of translated copies of this localized state over the entire lattice.}
\end{figure}

\begin{figure}
\centering\includegraphics[width=0.6\columnwidth]{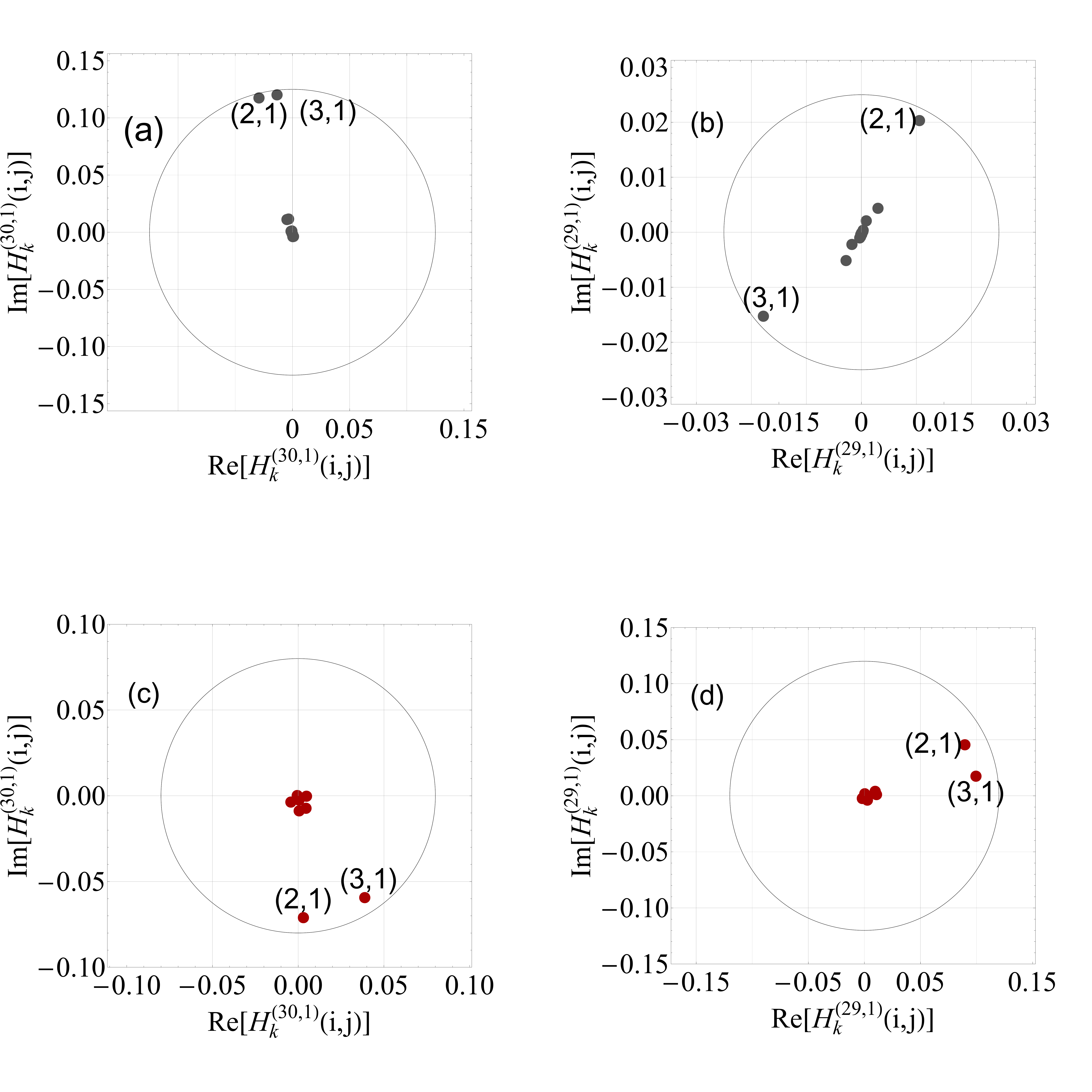}
\caption{\label{fig:PolarPlot} Matrix elements between the ground state ($m_i=1$) and the two consecutive excited states are shown in the complex plane for two quasi-momenta as in Fig.~\ref{fig:MatrixElementsNZQM}. On all sub-plots the two dominant matrix elements are marked by their corresponding indices of the equivalence classes. (a)-(b) For $q_3$ the two matrix elements are nearly the same and they are summed up to a significant value for (a) $m_f=30$ but nearly cancel each other due to opposite signs for (b) $m_f=29$. In the latter case the final state is of the opposite \textit{effective mirroring} to the ground state. (c)-(d) For $q_1$ the two matrix elements are always summed up to a large value with small variations between the final state (c) $m_f=29$ and (d) $m_f=30$. The axes signify real and imaginary parts of matrix elements $H_k^{(m_f,m_i)}(i,j)\equiv{C_i^{(m_f)}}^*C_j^{(m_i)}\langle\xi_n^{(i)}|\hat{H}_k|\xi_n^{(j)}\rangle$.}
\end{figure}

The selection rule proved in the previous section cannot be extended to $q_n\neq 0$ sub-spaces where Eq.~(\ref{Eq:MTCommutate}) becomes invalid. 
However, in this case we still observe significant variations in the numerical values of transition amplitudes which follow the general pattern discussed in the previous section (see Fig.~\ref{fig:MatrixElementsNZQM}).
For some quasi-momenta the contrast of variations is small (red circles), but for others it can reach as large values as an order of magnitude (gray squares).
This behavior is intimately related to the special form of the Hamiltonian's~(\ref{Eq:Hamiltonian}) eigenstates in the limit of $|U/J|\gg 1$. 
As we already mentioned in Sec.~\ref{sec:BHMSH}, the ground state is a localized state (see Fig.~\ref{fig:SolitonicState}) with nearly unity occupation of a single site~\cite{Sorensen12}. 
Note that in the mean-field limit it corresponds to the translational symmetry-breaking bright soliton solution~\cite{PS16}.

It can be intuitively understood that for such a localized ground state the main contribution to the expansion~(\ref{Eq:EigenStateExpansion}) comes from a single equivalence class, namely $\mathcal{E}^{(1)}=\{|N,0,...,0\rangle,|0,N,...,0\rangle,...\}$, while the other classes are weighted with vanishingly small coefficients.
Then, the ground state expansion can be approximated by a single term:
\begin{equation}
\label{Eq:ApproxInitialState}
|\Psi_n^{(m=1)}\rangle \approx C_1^{(m=1)} |\xi_n^{(1)}\rangle.
\end{equation}  

When $\hat{H}_k$ is applied to this state, only two equivalence classes are involved in the resulting state decomposition, namely $\mathcal{E}^{(2)}=\{|N-1,1,...,0\rangle,|0,N-1,1,...,0\rangle,...\}$ and $\mathcal{E}^{(3)}=\{|1,N-1,...,0\rangle,|0,1,N-1,...,0\rangle,...\}$:

\begin{equation}
\label{Eq:PerturbedInitialState}
\hat{H}_k|\Psi_n^{(m=1)}\rangle =  -J\left(C_2^{(m=1)} |\xi_n^{(2)}\rangle + C_3^{(m=1)} |\xi_n^{(3)}\rangle\right),
\end{equation}
and only these two terms are expected to contribute significantly to the transition amplitude.
Note that $\mathcal{E}^{(2)}$ and $\mathcal{E}^{(3)}$ are {\it mirrored} equivalence classes and, in case of zero quasi-momentum, $C_2^{(m)}$ and $C_3^{(m)}$ cancel each other exactly when transition amplitude between the opposite mirroring states is considered (see Sec.~\ref{sec:MirroringProperty}).
This is not the case for general $q_n$, for which $C^{(m)}_j$ coefficients are imaginary and the eigenstates of the Hamiltonian are not the eigenstates of the mirror operator $\hat{\mathcal{M}}$.
However, numerics show that $C_2^{(m)}$ and $C_3^{(m)}$ are still nearly opposite when transition amplitudes between the ground state and either odd or even states from the first excited band are considered.
Other parameters of the system, such as quasi-momentum $q_n$, size of the system $M$ and interaction strength $U/J$ define which states (even or odd) satisfy this condition.
This dependence is studied below. 
For now we note that this remarkable \textit{near cancellation} causes the general pattern shown in Fig.~\ref{fig:MatrixElementsNZQM} to remain similar to the $q_0$ case (see Fig.~\ref{fig:MatrixElements}(b)).
Thus, we attribute the \textit{effective mirroring} property to each state for convenience.
Then, if the matrix element between the two states is small, we call them states with the opposite \textit{effective mirroring}.

In Fig.~\ref{fig:PolarPlot}(a)-(d) all matrix elements corresponding to two consecutive final states and two different quasi-momenta are shown in complex plane. 
The axes signify real and imaginary parts of individual matrix elements $H_k^{(m_f,m_i)}(i,j)\equiv{C_i^{(m_f)}}^*C_j^{(m_i)}\langle\xi_n^{(i)}|\hat{H}_k|\xi_n^{(j)}\rangle$ for all possible pairs of the equivalence classes $(i,j)$.
The sum over all such pairs gives a transition amplitude between the initial state $m_{i}$ and the final state $m_{f}$ (see Figs.~\ref{fig:MatrixElements}$\&$\ref{fig:MatrixElementsNZQM}).
In Fig.~\ref{fig:PolarPlot}(a)-(d)  we observe that there are two dominant matrix elements, while all the others are close to zero.
These elements are marked with the corresponding indices of the equivalence classes.
Fig.~\ref{fig:PolarPlot}(a),(b) correspond to $q_3$ and to the transition between the ground ($m_i=1$) and the excited (a) $m_f=30$ and (b) $m_f=29$ states. 
In both cases the matrix elements are nearly the same but in the subplot (b) they are of the opposite signs and thus nearly cancel each other which results in the large contrast shown in Fig.~\ref{fig:MatrixElementsNZQM} as gray squares.
Fig.~\ref{fig:PolarPlot}(c),(d) correspond to $q_1$ and the same pair of excited states: (c) $m_f=30$ and (d) $m_f=29$.
In both case no significant difference is observed and the resulting contrast shown in Fig.~\ref{fig:MatrixElementsNZQM} in red circles remains small.
In Fig.~\ref{fig:PolarPlot}(b) it is clearly seen that the two main contributions are opposite in sign and, thus, they nearly cancel in the calculation of the matrix element.

\begin{figure}
\centering\includegraphics[width=0.6\columnwidth]{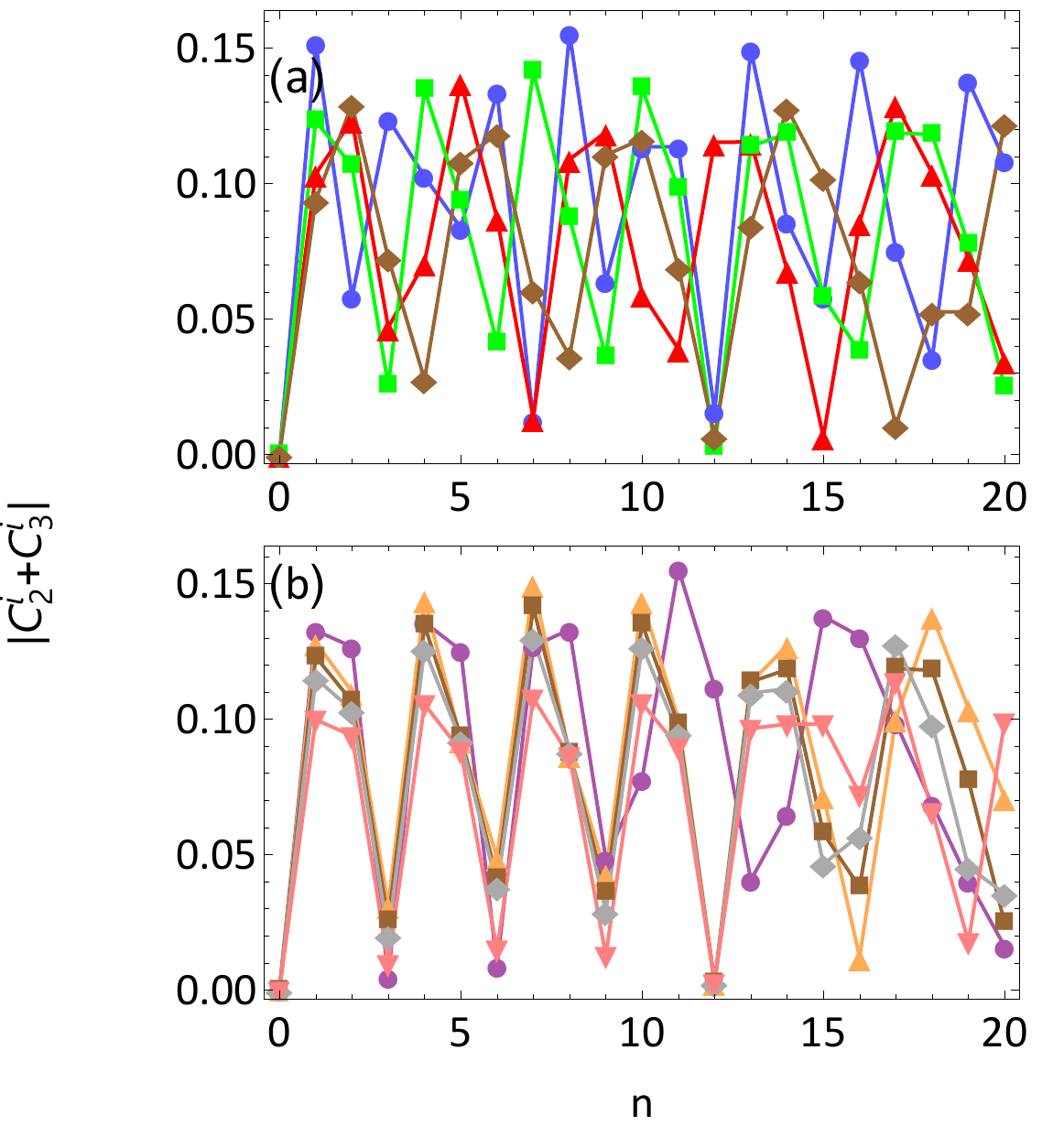}
\caption{\label{fig:MatrixElementsChaos} Irregular behavior of transition amplitudes as a function of quasi-momentum $q_n$ (X-axis represents index $n$ of the quasi-momentum). The Y-axis represents the contrast between the values of the maximal transition amplitude and its neighbour. Note that only contributions of two most significant matrix elements are included in the calculation of the transition amplitude. (a) The size of the system is fixed ($M=91$) and different interaction strengths are shown: $U/J=-7$ (blue circles), $U/J=-10$ (green squares), $U/J=-13$ (red triangles) and $U/J=-15$ (brown rhombus). (b) The interaction energy is fixed to $U/J=-10$ and different sizes of the system are shown: $M=74$ (purple circles), $M=83$ (orange up triangles), $M=91$ (brown squares), $M=101$ (gray rhombus) and $M=131$ (pink down triangles). }
\end{figure}

We now study this behavior for a wide range of quasi-momenta.
Fig.~\ref{fig:MatrixElementsChaos}(a) (Fig.~\ref{fig:MatrixElementsChaos}(b)) shows the most significant contributions to the transition amplitudes as a function of quasi-momentum for different $U/J$ and a fixed system size $M=91$ (for different $M$ and a fixed $U/J=-10$).
For small quasi-momenta some order can still be identified.
This is especially clear in Fig.~\ref{fig:MatrixElementsChaos}(b) where small coefficients are observed for $q_{n=3}$, $q_{n=6}$ and $q_{n=9}$ for all system sizes. 
For higher quasi-momenta this correlation is lost.
In Fig.~\ref{fig:MatrixElementsChaos}(a) the correlated minima does not exist.
In general, positions of small matrix elements are largely unpredictable and they appear rather irregularly as a function of changed parameter. 
This behavior might not be surprising as matrix elements result from the diagonalization of a large matrix and, thus, they are roots of a highly non-linear equation.
However, this fact deserves further consideration and will be the subject of future research.

\section{Induced transitions}
\label{sec:InducedTransitions}
In this section we solve the time dependent problem by direct integration of the time dependent Schr{\"o}dinger equation:
\begin{equation}
i\hbar\frac{\partial}{\partial t} |\Psi_n^{(m)}(t)\rangle = \hat{H}(t)|\Psi_n^{(m)}(t)\rangle,
\label{Eq:TDSE}
\end{equation} 
with the initial conditions of being in the ground state at $t=0$.
Projecting the Schr{\"o}dinger equation on different eigenstates we obtain a set of coupled differential equations for the time dependent occupation probabilities (see~\ref{sec:AppendixA}).

\begin{figure}
\centering\includegraphics[width=0.6\columnwidth]{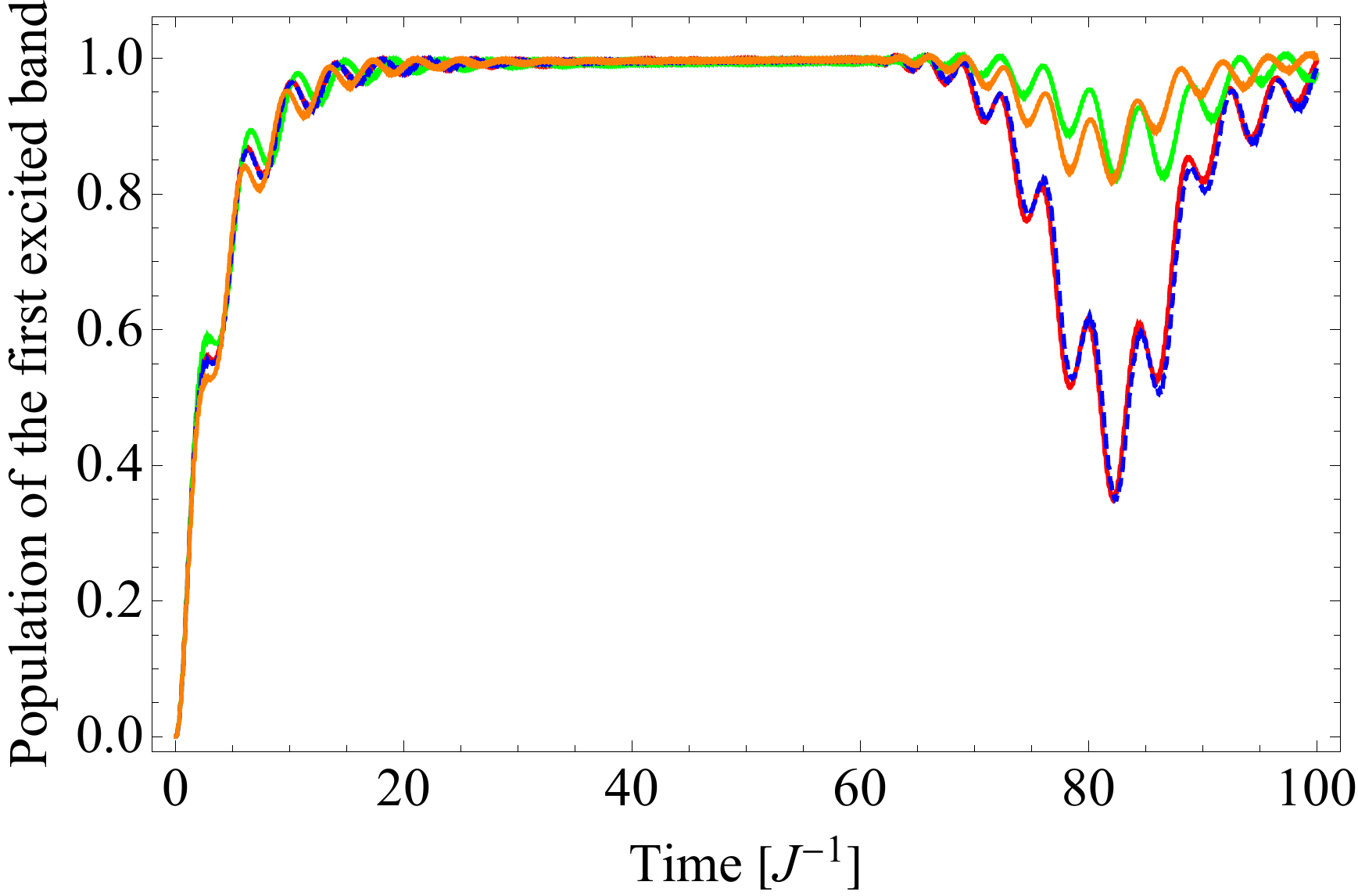}
\caption{\label{fig:OccupationProbability3atoms} Total occupation probability of the first excited band as a function of time for $N=3$, $M=149$ and $U/J=-10$. The four curves are for $q_0=0$ (dashed blue), $q_{n=1}$ (dark yellow), $q_{n=5}$ (green) and $q_{n=6}$ (red) quasi-momenta. All four curves are indistinguishable in the beginning but become very different when the population returns to the ground state at $\sim 75 J^{-1}$.}
\end{figure}

\subsection{Extracting one atom from a 3-atom soliton}
In Fig.~\ref{fig:OccupationProbability3atoms} we show the time dependent occupation probability of the first excited band which is the sum of the probabilities over all energy levels belonging to the same band.
Here we consider a system of $N=3$ atoms on $M=149$ cites whose energy spectrum is represented in Fig.~\ref{fig:EnergySpectrum} and the sum is performed over the green (central band) energy levels.
Solid lines describe the occupation probability of the first excited band for four different quasi-momenta: $q_0=0$ (blue), $q_{n=1}$ (dark yellow), $q_{n=5}$ (red) and $q_{n=6}$ (green).
After a few tens of tunnelling time, the population is totally transferred to the first excited band which corresponds to a two-atom localized state and one free atom.
Note that the occupation of the third band (corresponding to a complete disintegration of the three-atom localized state to three free atoms) remains always negligible.
This fact is directly reflected in numerical values of transition amplitudes for second (Fig.~\ref{fig:MatrixElements}(b)) and third (Fig.~\ref{fig:MatrixElements}(c)) bands respectively in case of strong interactions ($U/J=-10$).
As can be easily identified, the maximal value of the transition amplitude directly to the third band is suppressed by more than a factor of $30$.
In contrast, if weaker interactions are considered, transition amplitudes decay slowly for higher energy levels as shown in Fig.~\ref{fig:MatrixElements}(a) for $U/J = -3$.
In fact, in the latter case there is no band gap between the second and the third excited bands and the three-atom localized state can be directly disintegrated by a weak modulation.
This is, of course, a consequence of the finite kinetic energy associated with a periodic potential.
If the interaction is too weak to protect the localized states by a gap from further excitations, the soliton is disintegrated.

It is worth noting that the soliton exists in free space, i.e. when the strength of the periodic potential vanishes. In this limit no finite kinetic energy is associated with the periodic potential and exciting a single atom without destroying the localized state even for weak interactions is plausible.
The resonant modulation frequency (see Eq.~\ref{Eq:EnergyDifference}) in this case simply reduces to the chemical potential of the 1D attractive Bose gase.

In Fig.~\ref{fig:OccupationProbability3atoms}, at $t>60 J^{-1}$ the occupation probability of the first excited band decreases again and a minimum occurs at $\sim 75 J^{-1}$. 
In this minimum the population goes back to the ground state and this revival is expected due to coherent time evolution of the finite size system with nearly equally spaced energy levels.
The interesting feature of this revival is its contrast, which is significantly better in the case of $q_0=0$ and $q_{6}$ quasi-momenta (dashed blue and solid red lines in Fig.~\ref{fig:OccupationProbability3atoms}). 
This difference is a direct consequence of the selection (quasi-selection) rules dictated by mirror symmetry which vanishes (suppresses) transitions between the energy levels with opposite mirroring (effective mirroring). 
Therefore, for $q_0$ and $q_6$ quasi-momenta, only every second energy level in the excited band is involved in the dynamics effectively decreasing the dephasing rate and supporting a stronger revival.
For $q_1$  and $q_5$ (dark yellow and green lines in Fig.~\ref{fig:OccupationProbability3atoms}), in contrast, all energy levels participate in the transition which causes the dephasing rate to increase and the revival contrast to degrade.
We now note that the quasi-selection rules observed for other quasi-momenta can be identified in the revival contrast as well.
We numerically verified that the correlation between the appearance of quasi-selection rule and strong revivals holds for other quasi-momenta as well.
These revivals might serve as an experimental observable to detect quasi-selection rules.

\begin{figure}
\centering\includegraphics[width=0.55\columnwidth]{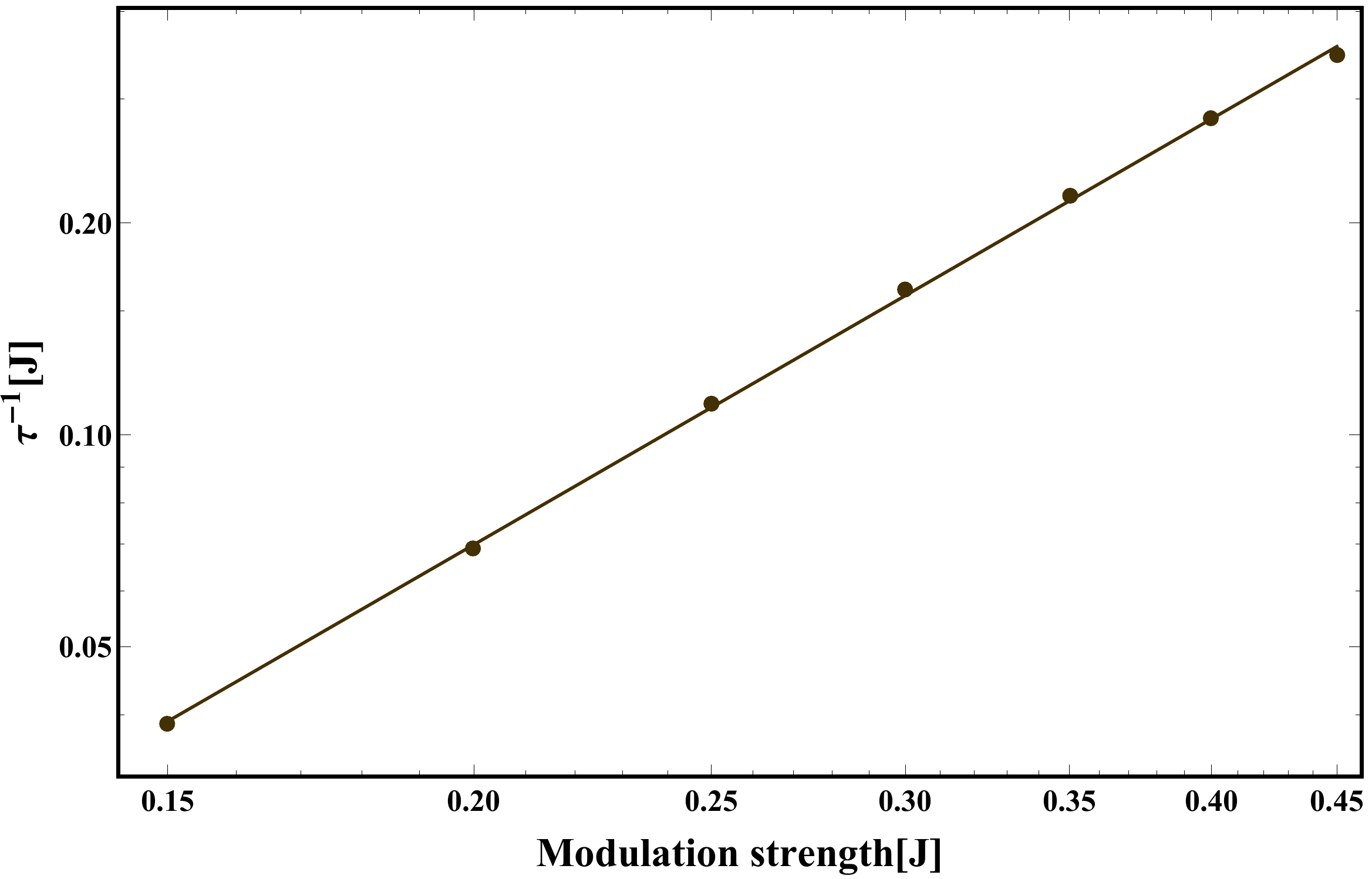}
\caption{\label{fig:FermiGoldenRule} Transition rate to the first excited band as a function of the modulation strength $\varepsilon J$ in log-log scale for $N=3$, $M=149$ and $U/J=-10$. The straight line is the fit to a power law function. The obtained slope ($2.07\pm0.04$) is in good agreement with the Fermi Golden rule.}
\end{figure}

For the initial-time dynamics the occupation of the first excited band increases exponentially and we verify that the transition rate associated with it follows the Fermi Golden rule quite precisely.
In Fig.~\ref{fig:FermiGoldenRule} (presented in log-log scale) we show that the transition rate increases as a square function of the modulation strength.
The solid line represents a fit to the numerical data and the obtained slope $2.07\pm0.04$ is in good agreement with the Fermi Golden rule.

\subsection{Extracting one atom from a 4-atom soliton}
\begin{figure}
\centering\includegraphics[width=0.6\columnwidth]{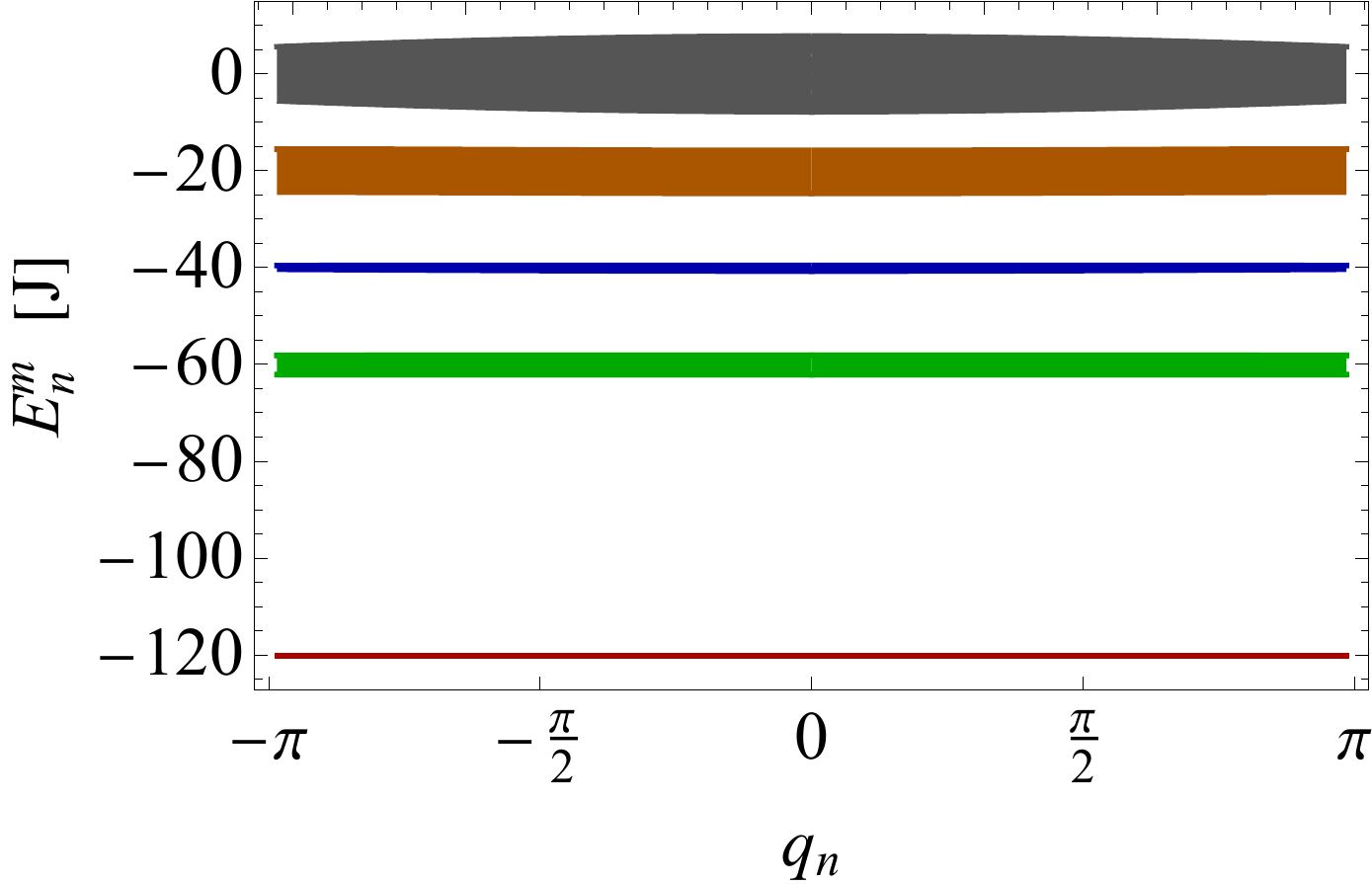}
\caption{\label{fig:EnergySpectrum4atoms} Energy diagram of $N=4$ particles on $M=65$ sites for $U/J=-20$. The ground state (red line) corresponds to all $4$ atoms in the localized state. The first excited band (green) describes $3$ atoms in the localized state and $N_f=1$ free atom. The second band (blue) describes the two $2$-atom solitons and the third excited band (brown) corresponds to $2$ atoms in the localized state and $N_f=2$ free atoms. The last excited band corresponds to $N_f=4$ free atoms.}
\end{figure}

To explore the possibility of cascading extraction of single atoms out of a large solitonic state we consider a system with $N=4$ atoms. 
A band structure of $N=4$ atoms on $M=65$ sites is represented in Fig.~\ref{fig:EnergySpectrum4atoms} for $U/J=-20$.
A larger interaction strength is needed in this case to preserve the finite band gap between the second and higher excited bands.
The energy spectrum now is split into a ground state and four excited bands. 
As before, the ground state corresponds to the CoM motion of the solitonic state of all 4 atoms.
The first excited band describes the 3 atom solitonic state and different translational modes of a free atom.
The next band corresponds to the relative motion of two localized sates each composed of 2 atoms.
The third and the forth excited bands describe a two-atom solitonic state and two free-atoms and fully disintegrated (4 free atoms) state respectively.
It is worth noting that the energy difference between the ground state and the first excited band coincides with the difference between the first and the last excited bands.
This coincidence affects the time evolution as we discuss below.

\begin{figure}
\centering\includegraphics[width=0.6\columnwidth]{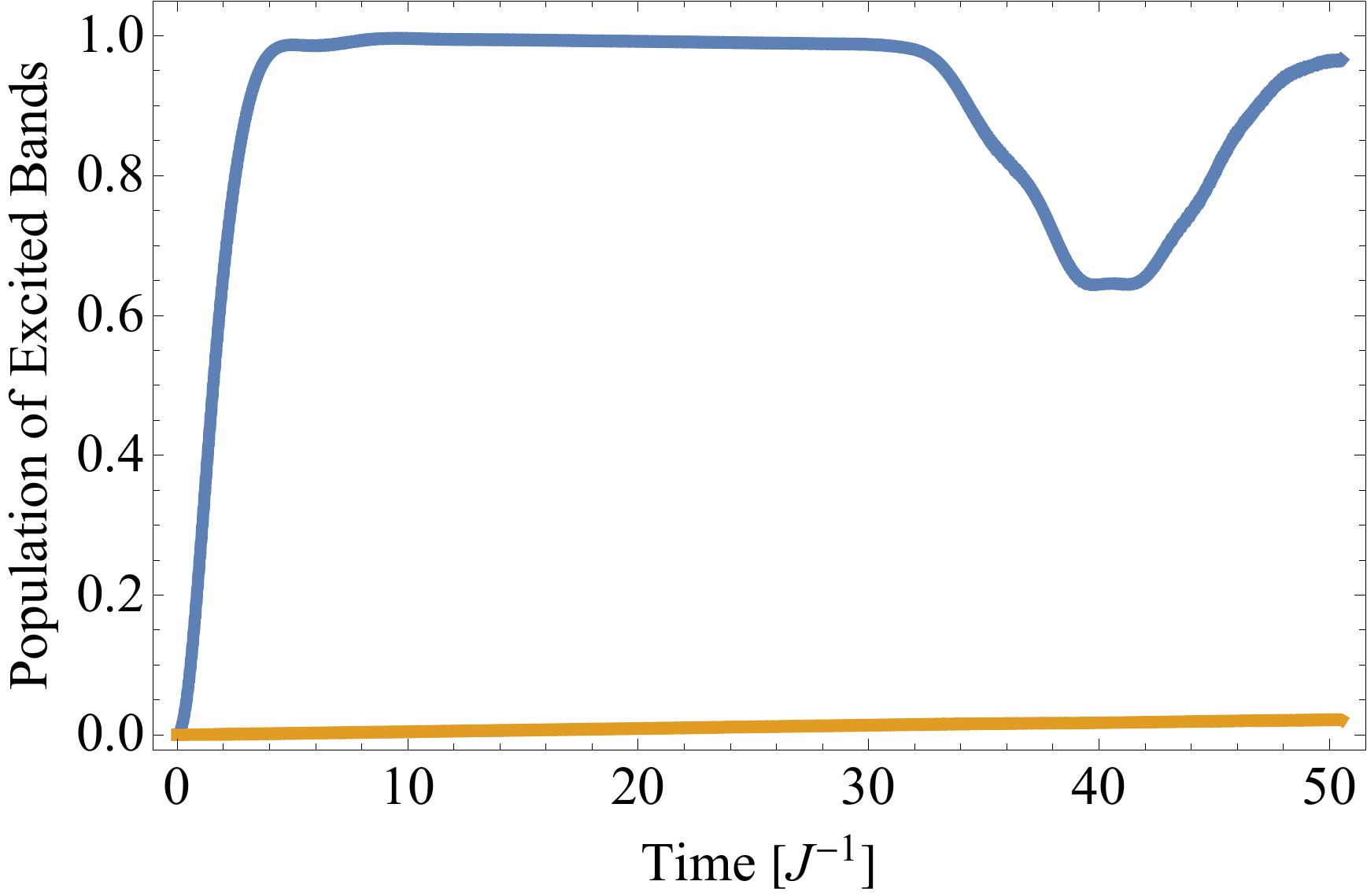}
\caption{\label{fig:OccupationProbability4atoms} Total occupation probability of the first excited band (sum over all but first excited bands) as a function of time for $N=4$, $M=65$ and $U/J=-20$ and zero quasi-momentum is shown by blue (orange) line. After a few tunnelling times the population is transferred to the first excited band while the population of other excited bands remains vanishingly small for short times.} 
\end{figure}

As before, we solve a set of coupled differential equations derived from the time dependent Schr{\"o}dinger equation Eq.~(\ref{Eq:TDSE}) restricting our analysis to the case of zero quasi-momentum. 
The sum of time dependent occupation probabilities of all states belonging to the first excited band is shown in Fig.~\ref{fig:OccupationProbability4atoms} as a blue line.
The orange line represent the sum over states belonging to all other excitation bands.
The observed behavior is qualitatively similar to the previously discussed case of $N=3$ atoms (Fig.~\ref{fig:OccupationProbability3atoms}). 
Most importantly, after a few tunnelling times the population is fully transferred to the first excited band which corresponds to a 3-atom localized state and one free atom. 
Population of other excited bands remain negligible for short times but for longer time evolution they become populated.
Mainly, the population grows in the highest excited band due to the above-mentioned coincidence between the energy differences (see Fig.~\ref{fig:EnergySpectrum4atoms}) which preserves the same resonance condition as required to excite a single atom out of a soliton.
However, this excitation is of second order and can be kept small for weak modulation amplitudes. 
For example, the population of the last band remains below 1$\%$ at the time of the first revival when the parameters of Fig.~\ref{fig:OccupationProbability4atoms} are used.

This calculation demonstrates the idea of cascading extraction of single atoms from an initially localized state.
When a 4-atom soliton is prepared, on-resonance modulation can be applied to extract one atom out of it.
Continuous modulation at the same frequency causes no damage to the remaining 3-atom soliton.
In order to remove one more atom, the modulation frequency has to be tuned to the new resonance condition.
This scheme can be efficient only for a small number of atoms in the localized state for which the relative change in binding energy is significant when a single atom is removed.
Note, that the total interaction energy of the localized state is proportional to $N^3$~\cite{Gertjerenken12}. 

\subsection{Experimental considerations}
Small systems of ultracold atoms trapped in highly controlled periodic potentials have been recently demonstrated in experiments~\cite{Barredo16,Endres16,Kim16}, and may provide a platform for realization of a system with a small number of attractively interacting atoms, trapped in a ring-shaped one dimensional optical lattice.
A modulation of the tunneling rate is readily obtained by weak modulation of the optical lattice's amplitude, a technique already introduced in the experiments~\cite{Stoferle04,Iucci06,Kollath06}. 
In such a configuration selection rules can be verified through the detection of the revival contrast while a well defined initial quasi-momentum can be prepared by means of a Doppler sensitive two-photon Raman transition.
The latter allows preparation of the initial wavepacket with a sub-recoil energy resolution~\cite{Moler92}.  

However, the demonstration of cascading extraction of a single atom from a localized solitonic state does not require the presence of the optical lattice.
It can be demonstrated in a quasi one-dimensional wave guide without the superimposed periodic potential.
In this configuration the modulation parameter has to be the interaction strength and the modulation frequency should be tuned to the continuum threshold (according to equation Eq.~(\ref{Eq:EnergyDifference})).
This free space realization is best approximated by the strongly interacting BHM ($|U/J|\gg 1$), because the finite kinetic energy associated with the optical lattice tends to destroy the weakly bound localized states if they are not protected by strong enough attractive interactions.

\section{Conclusion}
\label{sec:Conclusion}
We study induced transitions in the attractive BHM with periodic boundary conditions in the limit of strong interactions.
Transitions are excited by on-resonance modulation of the tunnelling rate.

We study selection rules that apply to the system and show that, apart from an obvious selection rule related to the translation invariance of the system, there is a sub-space of the total Hilbert space where an additional rule applies.
We identify a mirror symmetry in the zero quasi-momentum sub-space which dictates this rule.
Although it is strictly applicable exclusively to the $q_0=0$ sub-space, the specific structure of the eigenstates of the problem extends the applicability of this selection rule and dictates a complex structure of quasi-selection rules for arbitrary quasi-momenta.
Finally, we note that identified selection and quasi-selection rules are applicable for larger systems as well (i.e. for $N>3$).

We show that a single atom can be extracted out of a localized 4- or 3-atom state while leaving the 3- or 2-atom localized state untouched.
Direct extension of the model suggests the possibility of cascade extraction of atoms out of a few-atom localized state one by one. 
The limit on the number of atoms for which this protocol works remains to be studied in future research.

\ack
We thank Emanuele Dalla Torre for several fruitful discussions and for critical reading of the manuscript.
We acknowledge discussions with K. M{\o}lmer at early stages of the project.
This research was supported by the Israel Science Foundation (Grant No. 1340/16) and FIRST Program (Grant No. 2298/16).

\appendix
\section{The time dependent occupation probabilities.}
\label{sec:AppendixA}

The most general form of the wavefunction that solves the time dependent Schr{\"o}dinger equation~(\ref{Eq:TDSE}) with the Hamiltonian~(\ref{Eq:TDHamiltonian}) is:
\begin{equation}
|\Psi^{(m)}_n(t)\rangle = e^{-i\omega_m t}\sum_{l=1}^S c_l(t) e^{-i(\omega_l - \omega_m) t} |\Psi^{(l)}_n\rangle,
\end{equation}
where $S$ is the number of states for each qasimomentum $q_n$.

Substituting this solution into Eq.~(\ref{Eq:TDSE}) and projecting it to an eigenstate $\langle \Psi^{(k)}_n|$ we obtain the differential equation for the probability amplitude of finding an atom in a state $|\Psi	^{(k)}_n\rangle$:
\begin{equation}
\frac{c_k(t)}{dt}e^{-i\omega_kt} = -i\varepsilon \sin(\omega t)\sum_{l=1}^S c_l(t) R_{kl} e^{-i \omega_l t},
\end{equation}
where $R_{kl}=\langle \Psi_n^{(k)}|\hat{H}_k|\Psi_n^{(l)}\rangle$. 
In total, there are $S$ coupled differential equations for each possible state $|\Psi^{(k)}_n\rangle$ which we solve numerically.

\section*{References}


\end{document}